# Superconducting NbRuB derived through the fragment formalism

Weiwei Xie, Huixia Luo, K. Baroudi, J. W. Krizan, B. F. Phelan and R. J. Cava

Department of Chemistry, Princeton University, Princeton, NJ, USA, 08540

*Abstract*

We employ the chemical fragment formalism to perform a targeted superconductor search in the Nb-Ru-B system, yielding the orthorhombic metal-rich boride NbRuB, which displays BCS-like superconductivity with a $T_c$ = 3.1 K. NbRuB is derived from the chemical fragments $Nb_3B_2$ + $Ru_3B$, in which the $Nb_3B_2$ fragment contains B-B dimers and the $Ru_3B$ fragment contains isolated B atoms. A charge transfer occurs between the fragments. The results indicate that the fragment formalism is a useful chemical tool for the design of new intermetallic superconductors much the same way as the charge reservoir concept has been a useful chemical tool for the design of new copper oxide superconductors.

## Introduction

Borides are an important class of non-molecular solids.[1-2] Due to the light mass of boron, borides can sometimes display high superconducting transition temperatures; $MgB_2$ ($T_c$ = 39 K) and $YB_6$ ($T_c$ = 7 K) are important examples.[3-10] Here, motivated by the fragment formalism widely used when viewing the structures of inorganic and organometallic molecules, we describe the discovery of a new superconductor made through manipulating the valence electron concentration in a boride – not through doping via atom substitutions, but by doping through the combination of electron-donating and electron-accepting structural fragments. [11]

We find that the metal-rich boride, NbRuB, whose existence and structure have also recently been described by others [12], displays BCS-like superconductivity with a $T_c$ = 3.1 K. The crystal structure of this material, shown in Figure 1, is built up by two types of boron-centered trigonal prisms: an uncapped boron-centered trigonal prism with the formula $Ru_3B$ whose structure is as found in $Re_3B$, a known superconductor with a $T_c$ of 4.8 K, and a B-B dimer-containing face-sharing double trigonal prism of formula $Nb_3B_2$. Given these fragments, and the electron count of the $Re_3B$ superconductor, we start with the hypothesis that in the optimal case for superconductivity the structural fragments in NbRuB would be combined to yield $(Ru_3B)^{3+}(Nb_3B_2)^{3-}$ (i.e. $Nb_3Ru_3B_3$). This is because $Ru_3B$ has 3 electrons in excess of $Re_3B$ and we wish its electron count to be equivalent to that of $Re_3B$ to favor superconductivity; the $(Ru_3B)^{3+}$ fragment, with isolated B atoms, would then be both isostructural and isoelectronic with $Re_3B$, and the $(Nb_3B_2)^{3-}$ fragment, with B-B dimers, would be isoelectronic with NbB ($Nb_3B_3$); the $Nb_3B_2$ fragment would thus be an electron acceptor, and the $Ru_3B$ fragment would be an electron donor. We show from electronic structure calculations that this is indeed the case, and although the degree of charge transfer is less than optimal, superconductivity is none-the-less observed. This type of donor-acceptor structural fragment analysis has been used to explain the relative stabilities of the $ThCr_2Si_2$ and $CaBe_2Ge_2$ structure types [13-14]; here we have extended its use to the search for new superconductors.

## Synthesis and crystal structure

The samples were synthesized by arc melting elemental starting materials in a water-cooled copper hearth. Niobium (99.9%, Alfa Aesar), ruthenium (99.95%, Aldrich), and boron (99.999%, J&M) were weighed in the NbRuB stoichiometric ratio with 10% molar excess B added in order to balance the light element B loss during the arc melting. The buttons were turned and melted several times to ensure good homogeneity. Weight losses during the melting process were less than 1%. Annealing the as-cast products below 1400 °C yielded $Nb_3Ru_5B_2$ as the dominant phase, indicating that the NbRuB compound is stable only at higher temperatures. The as-cast arc-melted NbRuB sample was examined by powder X-ray diffraction for identification and phase purity on a Rigaku powder diffractometer employing Cu Kα radiation with aid of a full-profile Rietveld refinement using LHPM RIETICA. [15-16] The major phase in the powder pattern was a good fit to the NbRuB structural model we obtained from our single crystal study (described below). The quantitative analysis of the powder diffraction pattern showed that the polycrystalline sample employed for the bulk property characterization consisted of 79(1)% NbRuB, 18(1)% $Nb_3Ru_5B_2$[17] and 3(1)% NbB [18] (See Figure 2.) For the purposes of property comparison, $Nb_3Ru_5B_2$ was prepared as a pure phase by arc melting Nb, Ru and B in a 3:5:2 ratio and annealing the product at 1400 °C for 48 hours.

To specify the structure of NbRuB, single crystals were investigated on a Bruker Apex II diffractometer with Mo $Kα_1$ radiation. The crystal structure was solved using direct methods and refined by SHELXTL. [19] This material has an orthorhombic structure with space group *Pmma*, as shown in Figure 1. It is a layered structure containing planes of Ru-B interleaved with planes of Nb-B, alternating along the *b* axis. The detailed crystallographic data is shown in Tables 1 and 2; it is in agreement with the data reported in reference 11.

**Calculation and electronic structure**

To gain further insights into the electronic character of NbRuB, TB-LMTO-ASA calculations [20] were carried out to evaluate and analyze the electronic density of states (DOS) and the band structure. Within the local density approximation (LDA) [21], the corresponding DOS curves and band structure for NbRuB are illustrated in Figure 3, which emphasizes contributions from the Nb and Ru valence orbitals. In the LDA DOS curve, the Nb + Ru 4*d* band exhibits little fine structure except for a noticeable pseudogap at approximately –0.75 eV and a broad, intense peak at 0 eV. According to the corresponding -COHP curves, this latter peak is strongly Ru-Ru and Ru-Nb antibonding in character. Thus, according to the LDA-DOS curves, due to its relatively high DOS at $E_F$ derived from antibonding interactions, NbRuB is electronically unstable with respect to a structural distortion, itinerant magnetism, or, possibly, superconductivity. Applying spin polarization via the local spin density approximation (LSDA) splits the DOS curves for the spin-up and spin-down wavefunctions, but the summed DOS curves in LSDA are the same as the DOS curve in LDA. Integration of the spin-up and spin-down DOS curves yields a total magnetic moment of ~0 $\mu_B$ per formula unit. Thus itinerant magnetism is not expected, and the electronic structure calculations therefore support the possibility for superconductivity in NbRuB.

In intermetallic compounds like these, without clear formal differences in element electronegativity, metallic characteristics may appear to exclude the possibility for charge transfer between atoms, but this does not discount the possible presence of differences between the orbital occupations of the atoms in the compound from those of their ground state neutral gaseous atoms. To quantify the charge transfer in this material, we have used Bader charge analysis [22] based on density functional theory calculations performed by VASP [23]. Using this method, we find that the electronic distribution in the $Nb_3B_2$ fragment, which came with 21 intrinsic electrons, yields 23.18 e-, (in other words this fragment is ~ $(Nb_3B_2)^{2-}$) and the electronic distribution in the $Ru_3B$ fragment, which came with 27 intrinsic electrons, yields 24.82 e-/f.u. (in other words this fragment is ~$(Ru_3B)^{2+}$). These electron distributions are consistent with what is obtained when calculating the integrated DOS of the $Nb_3B_2$ (22.06e-) and $Ru_3B$ (25.94e-) fragments obtained from the LDA calculations. Thus, as hypothesized, the $Nb_3B_2$ part of the structure has accepted electrons (between approximately 1 and 2 electrons depending on the calculation used) from the $Ru_3B$ part of the structure, while the $Ru_3B$ part has lost a corresponding number of electrons, showing that the fragment formalism is a valid way to consider the electron distribution in NbRuB. Although the calculated charge transfer is less than the amount hypothesized as optimal for superconductivity, a new superconductor indeed occurs as described below.

**Electronic Characterization**

The temperature ($T$) dependent electrical resistivity ($\rho$) of NbRuB in the vicinity of the superconducting transition, measured by a four-probe technique using silver paste electrodes in a Quantum Design Physical Property Measurement System (PPMS), is shown in Figure 4. The resistivity undergoes a drop to zero at 3.1 K, characteristic of superconductivity. In correspondence with $\rho(T)$, the magnetic susceptibility ($\chi_{mol}(T)$), measured in a field of 10 Oe after zero field cooling using a Quantum Design, Inc. Superconducting Quantum Interference Device (SQUID) magnetometer starts to decrease at 3.1 K and shows large negative values, characteristic of an essentially fully superconducting sample, on lowering the temperature. The zero resistivity and the large diamagnetic susceptibility indicate that NbRuB becomes a bulk superconductor at 3.1 K.

To prove that the superconductivity is intrinsic to NbRuB, and is not a consequence of the impurity phases present, the superconducting transition was characterized through heat capacity measurements. The heat capacity for NbRuB in the temperature range of 1.9 to 40 K is presented in Figure 5. The main panel shows the temperature dependence of the zero-field heat capacity $C_p$. The good quality of the sample and the bulk nature of the superconductivity are supported by the presence of a large anomaly in the heat capacity at $T_c$= 3.0 ~ 3.1 K in the plot of $C_p$/T, at a temperature that is in excellent agreement with the $T_c$ determined by ρ(T) and χ. The electronic contribution to the specific heat, γ, measured in a field of 5 T to suppress the superconductivity (inset to Figure 5), is 10 mJ/mol-K$^2$. The value of the specific heat jump at $T_c$ is thus found to be consistent with that expected for a weak-coupling BCS superconductor; $\Delta C_p/\gamma T_c$ per mole NbRuB in the 78% pure sample = 0.85.[24] This ratio is not at the BCS superconductivity weak coupling limit of 1.43 but is in the range observed for many superconductors.[25] The superconductivity property parameters are summarized in Table 2. As an added check, we tested pure $Nb_3Ru_5B_2$ (present at the 18% level in the tested sample) down to 0.4 K and found that it is not superconducting; that compound therefore could not give rise to the observed heat capacity feature. Finally NbB is reported to be either antiferromagnetic or superconducting depending on preparation method [26-27], but even if superconducting, at only 3% of the tested sample it could not possibly give rise to a heat capacity anomaly of the size observed ($\Delta C_p/\gamma T_c$ per mole NbB would be about 22, a physically impossible value). Thus the observed superconductivity originates from NbRuB.

**Conclusion**

In conclusion, we report the results of a directed search for superconducting borides through the use of the chemical fragment formalism. This search yielded NbRuB as a candidate superconducting material based on the donor-acceptor relationship of the structural fragments present and the understanding of the structure and electron count of the $Re_3B$ superconductor. Although they do not address the complexity of the root physical causes of the superconductivity in copper oxide and iron arsenide superconductors, the analogous chemically-based charge reservoir concept has been a useful in designing new superconducting compounds in those families[28-29]. The work described here shows that the fragment formalism, typically applied to molecules rather than solids, is a useful concept for the design of new non-molecular superconductors. Comparison between the fragment formalism and the charge reservoir concept is shown in Figure 6.

**Acknowledgement**

This research was supported by the Department of Energy, grant DE-FG02-98ER45706.

**Table 1. Single crystal crystallographic data for NbRuB at 296(2) K.**

| Nominal composition | NbRuB |
|---|---|
| Refined Formula | Nb$_3$Ru$_3$B$_3$ |
| F.W. (g/mol); | 614.37 |
| Space group; Z | *Pmma* (No.51); 2 |
| $a$ (Å) | 10.867(1) |
| $b$ (Å) | 3.1563(3) |
| $c$ (Å) | 6.3500(6) |
| $V$ (Å$^3$) | 217.67(4) |
| Absorption Correction | Multi-Scan |
| Extinction Coefficient | 0.0034(2) |
| µ(mm$^{-1}$) | 17.577 |
| θ range (deg) | 3.716-28.332 |
| *hkl* ranges | $-14 \leq h \leq 13$<br>$-4 \leq k \leq 4$<br>$-8 \leq l \leq 8$ |
| No. reflections; $R_{int}$ | 1329; 0.0116 |
| No. independent reflections | 336 |
| No. parameters | 33 |
| $R_1$; $wR_2$ (all *I*) | 0.0135; 0.0203 |
| Goodness of fit | 1.140 |
| Diffraction peak and hole (e$^-$/Å$^3$) | 0.760; −1.043 |

**Table 2. Atomic coordinates and equivalent isotropic displacement parameters of NbRuB** as refined from single crystal diffraction data. U$_{eq}$ is defined as one-third of the trace of the orthogonalized U$_{ij}$ tensor (Å$^2$).

| Atom | Wyckoff. | Occupancy. | $x$ | $y$ | $z$ | $U_{eq}$ |
|---|---|---|---|---|---|---|
| Ru1 | 2*f* | 1 | 0.3756(1) | ½ | 0.5724(1) | 0.0026(1) |
| Ru2 | 4*j* | 1 | ¼ | ½ | 0.2010(1) | 0.0041(1) |
| Nb1 | 2*e* | 1 | 0.5438(1) | 0 | 0.7772(1) | 0.0032(1) |
| Nb2 | 4*i* | 1 | ¼ | 0 | 0.8552(1) | 0.0025(1) |
| B1 | 2*e* | 1 | 0.4163(4) | ½ | 0.9499(7) | 0.0051(9) |
| B2 | 4*i* | 1 | ¼ | 0 | 0.450(1) | 0.008(1) |

**Table 3. Superconducting properties of NbRuB**

| | NbRuB | |
|---|---|---|
| $T_c$ (K) | -- | 3.1 |
| $\gamma$ (mJ/mol-K$^2$) $\beta$ (mJ/mol-K$^4$) | $C_p/T = \gamma + \beta T^2$ | $\gamma$= 9.97 $\beta$ =0.057 |
| $\theta_D$ (K) | $\Theta_D = (12\pi^4 nR/5\beta)^{1/3}$ | 468 |
| $\lambda_{ep}$ | $\lambda_{ep} = \dfrac{1.04+\mu^*\ln(\frac{\Theta_D}{1.45T_C})}{(1-0.62\mu^*)\ln(\frac{\Theta_D}{1.45T_C})-1.04}$ ($\mu^*$=0.15) | 0.544 |
| N($E_F$) experiment (states/eV NbRuB) | $N(E_F) = \dfrac{3}{\pi^2 k_B^2(1+\lambda_{ep})}\gamma$ | 2.74 |
| N($E_F$) calculation (states/eV NbRuB) | | 1.65 |
| $\Delta C/\gamma T_c$ | | 0.85 |

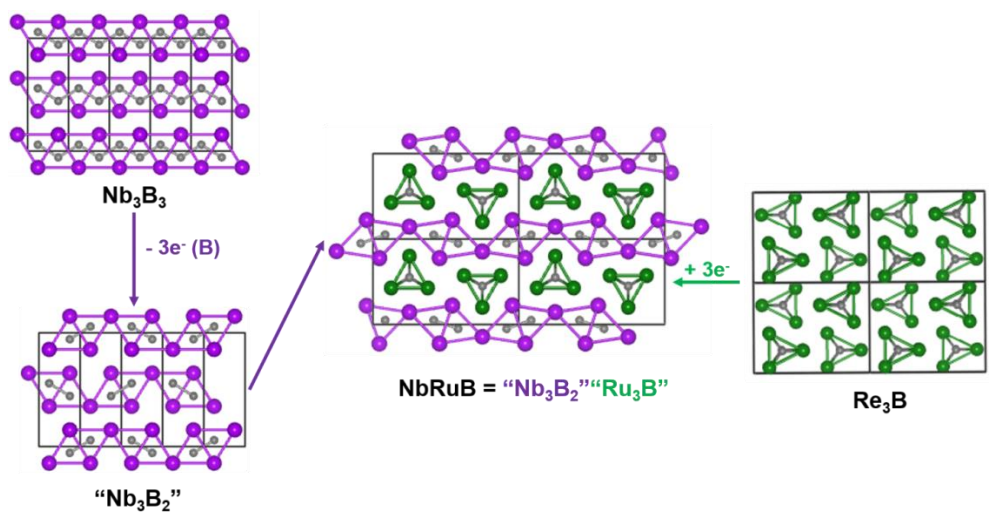

**Figure 1. Derivation of the crystal structure of NbRuB from structural fragments.** Center, the crystal structure of NbRuB determined from the single crystal refinement in a (010) view emphasizing the trigonal prisms B-B@Nb$_8$ and B@Ru$_6$. Left – the "Nb$_3$B$_2$" fragment is obtained after removing one B atom (3e-) from Nb$_3$B$_3$, whereas (right) the Ru$_3$B fragment is formed by adding 3e$^-$ to Re$_3$B. These fragments alternate in layers to create NbRuB.

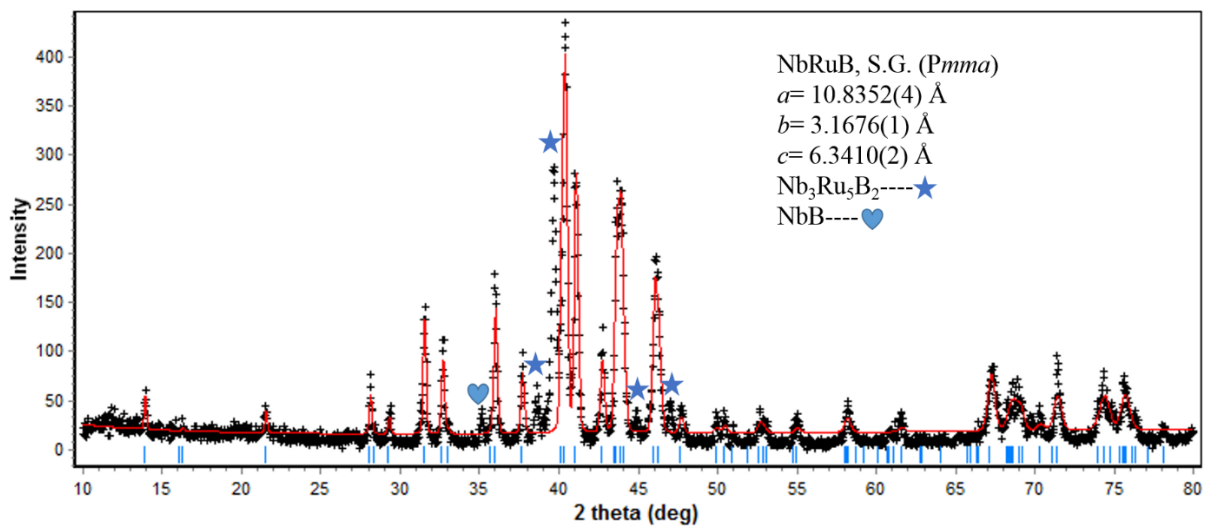

**Figure 2. Quantitative fit to the X-ray powder diffraction pattern for the sample employed for the property measurements** (Cu Kα radiation, 295 K, see text).

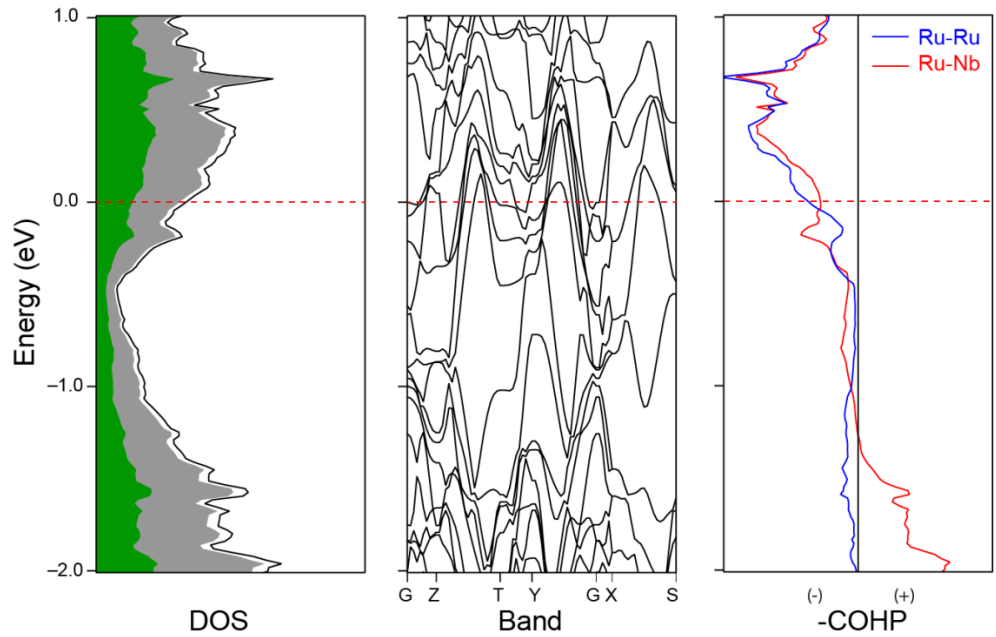

**Figure 3. Electronic density of states, band structure, and bonding/antibonding interactions for NbRuB.** From left to right, respectively: the partial DOS curves, the band structure curves, and the -COHP for NbRuB obtained from non-spin-polarized LDA calculations. (In the DOS curves, the Ru contribution is grey, and the Nb contribution is green). (In the −COHP: + is bonding/ − is anti-bonding, $E_F$ is set to zero.)

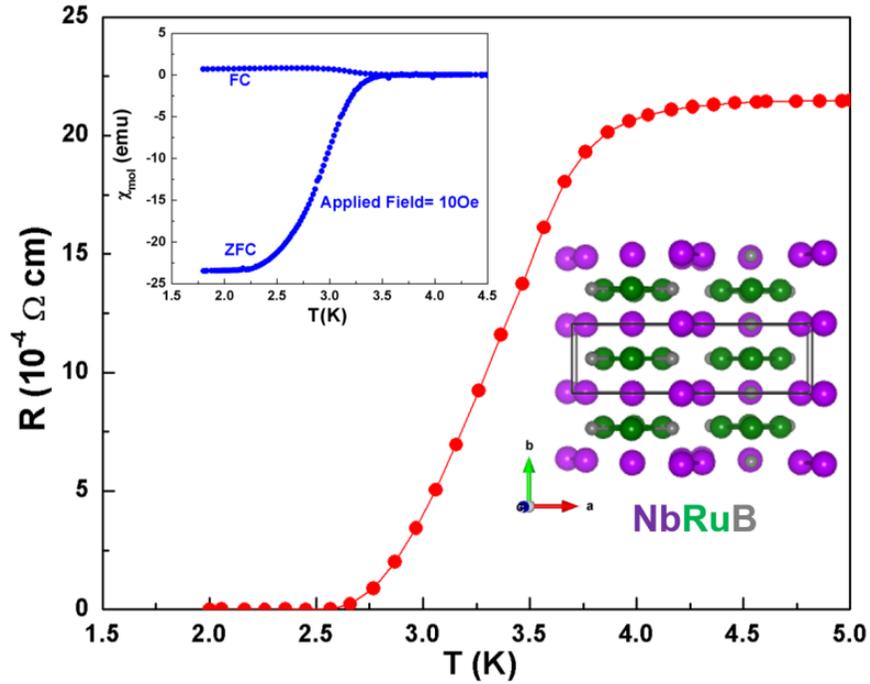

**Figure 4. The superconducting transition in NbRuB** (*Main panel*) The temperature dependence of the electrical resistivity of NbRuB without an applied magnetic field showing a close-up of the superconducting transition. Lower Insert: the crystal structure of NbRuB. Upper insert: the temperature dependence of the magnetic susceptibility of NbRuB in a 10 Oe applied field from 1.8K to 4.5K with zero-field cooling and field cooling.

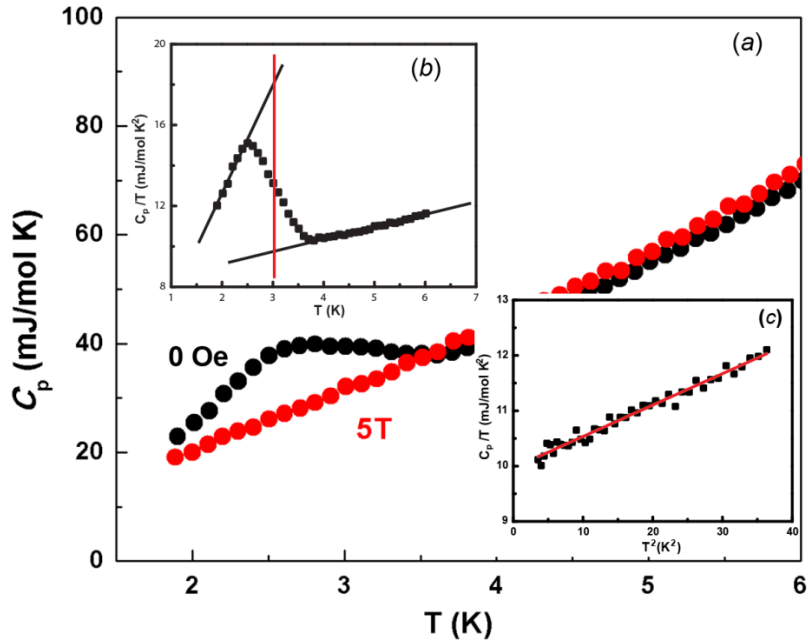

**Figure 5. Specific heat characterization of the superconducting transition of NbRuB.** (*a*) The temperature dependence of the heat capacity $C_p$ of NbRuB measured with (5T) and without an applied magnetic field (*b*) Enlarged view of the low temperature region (1.9 - 6 K) of $C_p/T$ (T) for NbRuB showing "the equal area construction" method for determining the change in entropy at the superconducting transition. (*c*) The fitting of the low temperature $C_p/T$ data vs. $T^2$ in the temperature range 1.9 - 6 K under the applied field of 5 T.

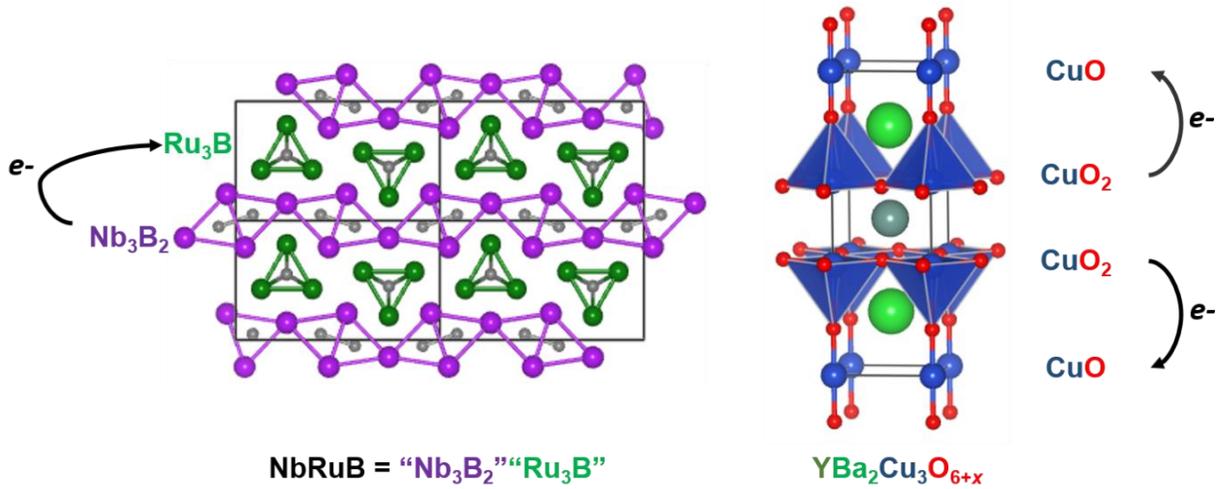

**Figure 6. Comparison of the crystal and electronic structures of the superconductors NbRuB and YBa$_2$Cu$_3$O$_{6+x}$ from the viewpoint of structural fragments, charge reservoir layers, and charge transfer.** Though charge transfer between intermediary layers and CuO$_2$ planes has been commonly used for discovering and understanding copper oxide superconductors, it has been rarely considered in the discovery of intermetallic superconductors. In this work we have used it to discover superconductivity near 3 K in NbRuB.